# Nanoparticle-doped electrospun fiber random lasers with spatially extended light modes

Vincenzo Resta[1], Andrea Camposeo[2], Martina Montinaro[1], Maria Moffa[2], Karolis Kazlauskas[3], Saulius Jursenas[3], Ausra Tomkeviciene[4], Juozas V. Grazulevicius[4], and Dario Pisignano[1,2,*]

[1]*Dipartimento di Matematica e Fisica "Ennio De Giorgi", Università del Salento, via Arnesano, I-73100 Lecce, Italy*
[2]*NEST, Istituto Nanoscienze-CNR, Piazza San Silvestro 12, I-56127 Pisa, Italy*
[3]*Institute of Applied Research, Vilnius University, Saulėtekio 3, LT-10257 Vilnius, Lithuania*
[4]*Department of Polymer Chemistry and Technology, Kaunas University of Technology, Radvilenu Plentas 19, LT-50254 Kaunas, Lithuania*
*\*dario.pisignano@unisalento.it*

**Abstract:** Complex assemblies of light-emitting polymer nanofibers with molecular materials exhibiting optical gain can lead to important advance to amorphous photonics and to random laser science and devices. In disordered mats of nanofibers, multiple scattering and waveguiding might interplay to determine localization or spreading of optical modes as well as correlation effects. Here we study electrospun fibers embedding a lasing fluorene-carbazole-fluorene molecule and doped with titania nanoparticles, which exhibit random lasing with sub-nm spectral width and threshold of about 9 mJ cm$^{-2}$ for the absorbed excitation fluence. We focus on the spatial and spectral behavior of optical modes in the disordered and non-woven networks, finding evidence for the presence of modes with very large spatial extent, up to the 100 μm-scale. These findings suggest emission coupling into integrated nanofiber transmission channels as effective mechanism for enhancing spectral selectivity in random lasers and correlations of light modes in the complex and disordered material.

**1. Introduction**

Random lasers are complex photonic devices which rely on multiple scattering of light in microscopically non-homogeneous media with optical gain [1]. When relevant interference effects are present for the scattered light, some degree of spatial localization can be predicted for optical modes, which reduces inter-mode coupling [2]. These phenomena have also been associated to the presence of very narrow (full width at half maximum, FWHM, down to sub-nm) lasing peaks [3]. Such sharp resonances are largely independent oscillations, typically observed in various systems which spontaneously embed random cavities, such as semiconductor powders [3,4], clusters of colloidal nanoparticles [2], and conjugated polymer films [5]. Methods to achieve so-called resonant feedback random lasing from independent modes include using excitation beams with high directionality, and designing weakly diffusive media as widely studied with solutions of lasing dyes with titania scatterers [2,6]. These aspects have motivated a debate about how much the modes of random lasers spread in the disordered medium, with various possible configurations proposed, ranging from highly localized states with exponentially decaying amplitude to extended modes related to scattering resonances [3,4,7-9].

Correlated emission and non-locality may also arise in random lasers, associated to mode competition and interplay through open cavities leading to extended enough and eventually overlapping light states [2,6,8,10]. In this case, the propagation of light is more likely to be describable as a diffusive process [6,11], with a characteristic photon transport mean free path, in analogy with random walking particles. The number of living modes drastically decreases upon increasing the strength of the interaction, until only cooperative modes survive and condensate into a single wavefunction. Such mode-locking regime is promising for applying these devices to several types of photonic chips, including platforms with all-optical control [2,10].

Random lasing has been achieved in a wide range of three-dimensional (3D) systems, largely using inorganic nanostructures and powders (e.g., ZnO [12,13], GaN [12], $BaSO_4$ [14] etc.), photonic glasses [15], or dispersions of elastic scatterers in solutions with lasing dyes [2,6,8,10]. Organic crystals [16,17] and epitaxial nanowires [18], biopolymers [19,20], as well as conjugated polymers [5,21], have also been found to efficiently work as random lasing materials. Affordable, large scale and easy processing procedures make organic building blocks, and their nanocomposites at the solid-state [13,22], especially sound for building low-cost and versatile devices, whose possible applications might span from speckle-free imaging schemes [23] to thermal [1], bio-chemical [24], and biological [25] sensing and diagnostics. In addition, some organic nanostructured systems, such as light-emitting polymer nanofibers [26], would allow unexplored aspects of mode condensation to be probed, since in quasi-one dimensional organic filaments a guided way of propagation for the lasing modes is added to diffusion and scattering.

In this work we analyze the spatial mode behavior in random lasing from disordered, non-woven networks of nanoparticle-doped polymer fibers based on a molecular active material with optical gain. This system undergoes lasing with very narrow emission peaks as typical of resonant feedback, and simultaneous presence of spatially extended modes, whose spreading might be supported through emitting regions connected by polymer nanofibers. Detailed space/spectrum cross-correlation (SSCC) experiments are carried out which highlight exceptional mode spreading up to 100 µm-scale and suggest coupling into fiber transmission channels as effective mechanism for enhancing correlations of the lasing modes in complex and disordered materials.

## 2. Experimental

*2.1 Materials and synthesis*

Fibers are electrospun from a 300 mg/mL chloroform solution of polystyrene (PS, molecular weight = 192,000 Da, Sigma-Aldrich) and an organic dye based on fluorene and carbazole triads, 2,7-Bis(9,9-diethylfluoren-2-yl)-9-(2-ethylhexyl)carbazole (Fl-Cz-Fl), at 5% w:w Fl-Cz-Fl:PS concentration. The synthesis details for Fl-Cz-Fl are reported elsewhere [27]. $TiO_2$ nanoparticles (Titanium(IV) oxide nanopowder, 21 nm average diameter, Sigma-Aldrich) are added at 1% w:w $TiO_2$:PS. Solutions are stirred for 12 h and inserted in a syringe with a 21 gauge metal needle. A flow of 1 mL/h is obtained by a peristaltic pump (33 Dual Syringe Pump, Harvard Apparatus Inc.) and a bias of 13 kV is applied between the needle and a metal collector at a distance of 10 cm, through a high-voltage power supply (EL60R0.6–22, Glassman High Voltage). Electrospinning is performed in air at ambient conditions, and fibers are collected on quartz substrates placed on top of the metal collector surface.

*2.2 Characterization techniques*

Fibers are inspected through optical microscopy (Olympus BX52 Microscope, Olympus Lifescience), and through scanning transmission electron microscopy (STEM, Nova NanoSEM 450, FEI) with 30 µm aperture size and 30 kV acceleration voltage. For determining the random lasing characteristics, the fibers are excited in vacuum (~ $10^{-1}$ mbar) by the 3$^{rd}$ harmonic (355 nm) of a Nd:YAG laser (Quanta-Ray INDI, Spectra-Physics) with about 7 ns pulse width (FWHM) at 10 Hz repetition rate, timing jitter of 0.5 ns and shot-to-shot energy variations of about 2%. *L-L* plots for the random lasers are obtained by shaping the pump beam, with Gaussian spatial profile, into a stripe (55 μm × 1.9 mm) through a cylindrical lens with 100 mm focal length, analyzing the emission from the sample edge by a monochromator (iHR320, JobinYvon) equipped with a Peltier-cooled charged-coupled-device (CCD) detector (Symphony, Horiba). The spectral resolution of the system is 0.15 nm.

The setup used for SSCC is schematized in Fig. 1. A 250 μm spot from the Nd:YAG laser is obtained by an iris pin-hole, a subsequent 3.3× expansion with a telescope system and a focalization onto fibers (L1 lens in Fig. 1). Part of the excitation is redirected (BS1) and monitored in real-time by a pyroelectric detector (J3-09, Coherent). The radiation emitted by fibers is imaged onto the entrance slit of the monochromator with a 5× magnification through an objective lens (L2) and a focusing lens (L3). The collected radiation is imaged by the Peltier-cooled CCD with a 1024×256 array of 26×26 μm$^2$, vertically-binned pixels allowing emission spectra to be collected for each coordinate along the direction defined by slit long axis. This leads to a resolution on the sample surface, parallel to the *Y*-direction, of about 10 μm ($Y_S$ coordinate, calculated by taking into account the magnification applied).

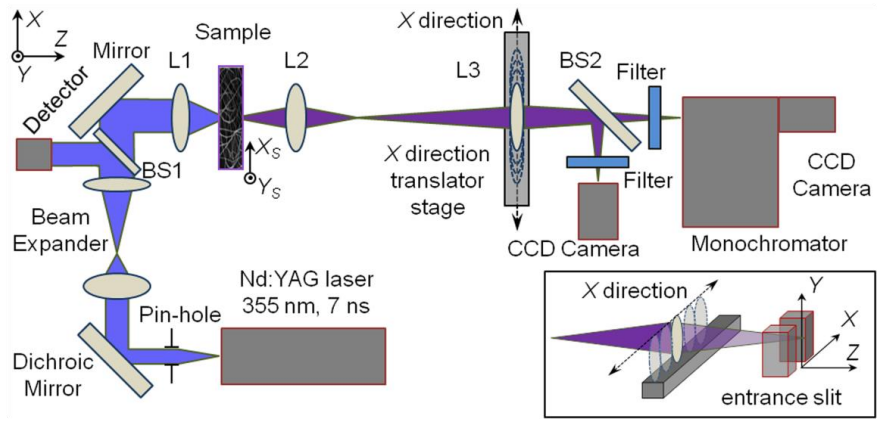

Fig. 1. Scheme of the setup for SSCC experiments, analysis of mode fluctuations and simultaneous fluorescence imaging. Samples are irradiated at normal incidence. The L3 lens (focal distance = 300 mm) is mounted on a micrometric linear translation stage for mode reconstruction along the direction ($X$) perpendicular to the slit long axis (bottom-right inset). Collections filters are longpass components removing the excitation line.

Moving the L3 lens along the $X$ direction perpendicular to the direction of the laser beam propagation and to the slit long axis ($Y$), with steps of 100 μm through a translation stage, leads to a resolution of 20 μm for the corresponding coordinate at the sample surface ($X_S$). The whole excitation field is reconstructed by the composition of 15 maps which are then shown by using contour lines as color-fill boundary. Image processing involves calculating a set of values by interpolating the slope (steepness) intensity between two adjacent pixels, which is in turn defined as the ratio of rise (difference in the intensity values) and run (pixel width). Spatial information is simultaneously acquired with far-field fluorescence imaging. To this aim a beam splitter (BS2 in Fig. 1) sends a part of the emitted radiation to a 405 nm longpass filter (Semrock) and to another CCD (1280×1024 pixel array, 5.2×5.2 μm$^2$ pixel size, DCC1545M, Thorlabs).

## 3. Results and Discussion

Our devices are built by disordered samples of electrospun PS fibers doped with a Fl-Cz-Fl. The molecular structure of Fl-Cz-Fl is shown in Fig. 2(a). Fl-Cz-Fl dispersed in a PS matrix features a broad absorption band at 3.54 eV, a fluorescence spectrum with clearly resolved vibronic bands from 3.12 eV to 2.77 eV, and a corresponding high quantum yield of 0.86 [28]. In the nanocomposite mat, Fl-Cz-Fl serves as gain material, whereas additional scattering can be provided in the system by doping with $TiO_2$ nanoparticles. Electrified Fl-Cz-Fl/PS solution jets biased at 13 kV are used to deposit fibers on quartz substrates, which are in turn placed onto the metal collector. The resulting morphology of a network of randomly oriented Fl-Cz-Fl/$TiO_2$-doped PS fibers is shown in Fig. 2(b).

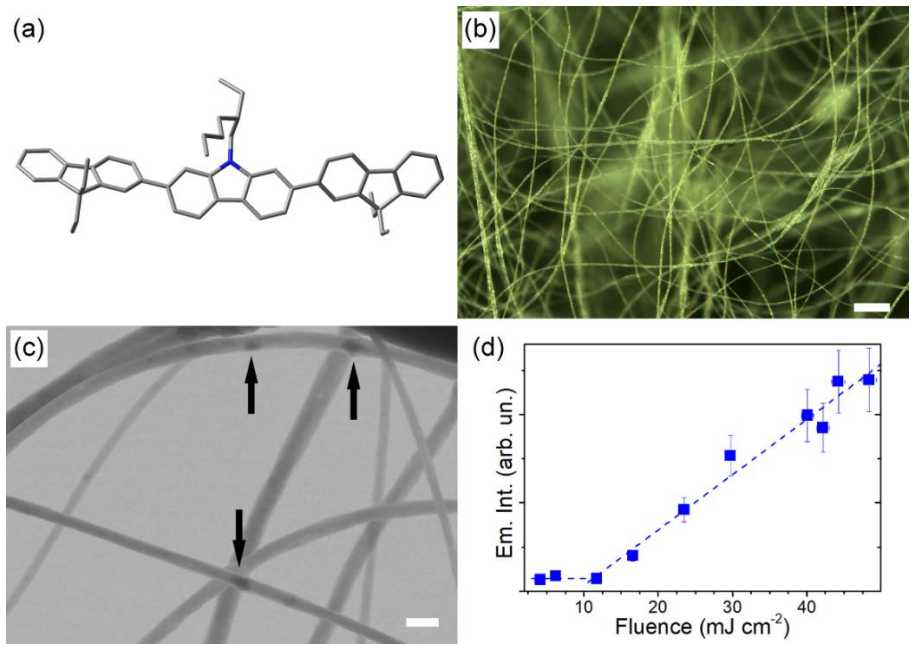

Fig. 2. (a) Molecular structure of Fl-Cz-Fl, geometrically optimized at the B3LYP/6-31G(d) level using density functional theory [27]. The blue segment indicates nitrogen atom. (b) Dark-field optical micrograph of electrospun, $TiO_2$/Fl-Cz-Fl-doped PS fibers. Scale bar: 50 µm. (c) STEM image showing $TiO_2$ clusters (dark spots highlighted by arrows) embedded in the fibers. Scale bar: 5 µm. (d) Random lasing emission intensity for the mode at 419.9 nm vs. excitation fluence. Intensity data are averaged over 10 excitation shots.

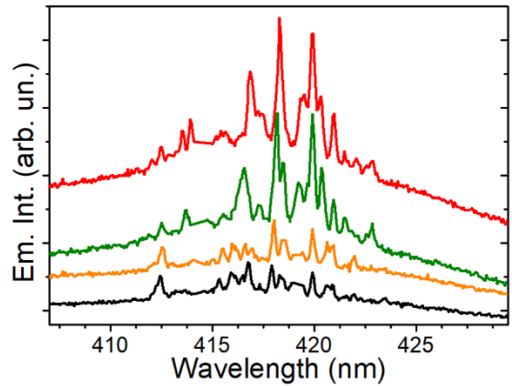

Fig. 3. Example of single-shot spectra measured with excitation fluence of 17 mJ cm$^{-2}$ (two bottom curves), and 42 mJ cm$^{-2}$ (two top curves), respectively.

The fibers have typical diameter ranging from 2 to 5 µm, and form a dense non-woven mat. In the dark-field micrograph in Fig. 2(b), with illumination from the top fibrous surface and signal being collected in back-scattered configuration, the different focusing conditions highlight the 3D character of the network, and light-scattering can be appreciated from the bright spots decorating fibers along their length, which is given by incorporated particles. These are better inspected by STEM, unveiling clusters distributed in the organic filaments, with mutual distances from a few to many tens of µm, as displayed in Fig. 2(c). Fig. 2(d) shows the *L-L* plot of the lasing emission (Light-out) vs. absorbed excitation fluence, delivered by a 55 µm-wide stripe at the third harmonic ($\lambda$=355 nm) of a 7 ns Nd:YAG laser (Light-in). The *L-L* characteristic is shown for the exemplary resonance occurring at the wavelength of 419.9 nm from Fl-Cz-Fl/TiO$_2$-doped fibers, revealing that the spectra undergo a rapid increase in intensity when the optical pumping exceeds a threshold of about 9 mJ cm$^{-2}$. At variance with Fabry-Pérot [29] or random-cavity deterministic lasers based on polymer nanofibers [30], where the lasing emission, originating from well-established cylindrical or ring-shaped cavities formed during fiber deposition, is characterized by well-defined and stable spectral features determined by the cavity geometry (i.e., fiber segment length or ring perimeter), here the overall number and intensity of the lasing modes is found to change from shot-to-shot, as displayed in Fig. 3 and in Fig. 4(a). In particular, in Fig. 3 we show two couples of single-shot emission spectra obtained with excitation fluence 17 mJ cm$^{-2}$ and 42 mJ cm$^{-2}$, respectively. The inter-mode shot-to-shot variability in the emission is not related to instabilities in the excitation fluence. Indeed, fluctuations of the pumping laser are below 2% (evaluated as the ratio between standard deviation and average value) in the time interval typical of spectroscopic measurements (a few minutes), whereas the measured shot-to-shot intensity variations of the fiber emission is in the range 5-20%, similarly to other random laser systems [31]. These fluctuations have been observed in various systems at the solid state, both in resonant-feedback [4,16,32], and in intensity-feedback random lasers [13], and attributed to either exceptional sensitivity of random lasers to the onset of spontaneous emission events, or the finite lifetime of the excited states of the gain material. Indeed, shot-to-shot variability of emission relates with the competition for gain, occurring each time in the whole and complex set of spatially distinct passive modes, established in the disordered material which is excited. Once activated, the lasing modes exhibit a spectral width as low as 0.15 nm (resolution-limited, Fig. 3). Ensemble averaging over an increasing number ($N$) of single-shot spectra (Figs. 4(b) and c) highlights a general smoothing of the emission spectrum, and stable average emission intensity from the lasing spectral region, for $N>10$. Furthermore, Fig. 4(d) shows the lasing behavior upon varying the excitation area.

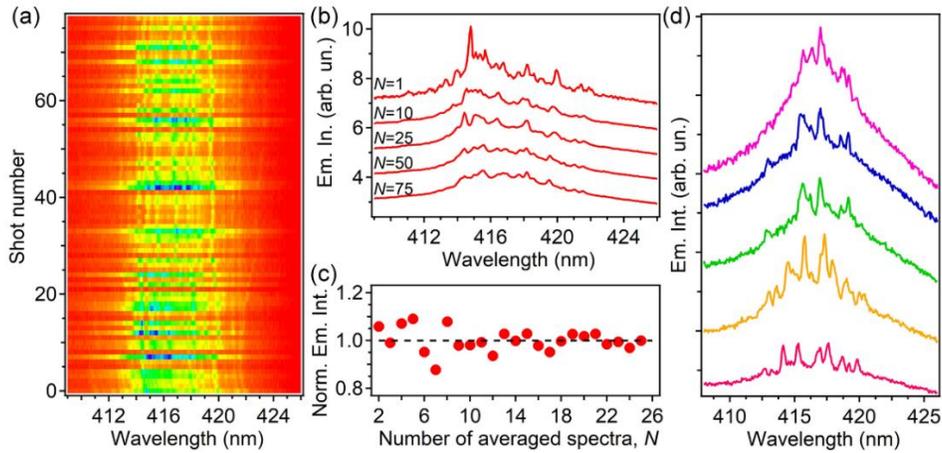

Fig. 4. (a) Evolution of the single-shot emission spectra upon increasing the number of excitation pulses. Excitation fluence 30 mJ cm$^{-2}$. (b) Averaged emission spectra calculated by using a varying number, *N*, of single-shot spectra. From top to bottom *N*=1, 10, 25, 50, 75, respectively. Spectra are shifted vertically for better clarity. (c) Dependence of the laser emission intensity on the number of single-shot spectra used for averaging. (d) Emission spectra collected for varied excitation areas. From bottom to top, the excitation area is 0.07, 0.13, 0.21, 0.33, and 0.47 mm$^2$, respectively. Excitation fluence 25 mJ cm$^{-2}$.

A broader and less spiky spectrum is found by increasing the excitation area, attributable to a larger number of activated lasing modes. In fact, controlling the spatial shape of the excitation beam constitutes an effective method for tailoring the spectral features of the random lasing emission, and for eventually selecting specific modes by active control [33].

The spatial distribution of modes, and the SSCC [5] is studied in our work by recording the dispersed spectrum by a CCD array, with resolution on the sample emission plane of ~10 μm/pixel, in binned configuration along the detector vertical direction (*Y*, parallel to the slit long axis, as detailed in the Experimental Section). A typical map of the spectrally- and $Y_S$-resolved emission is given in Fig. 5(a), showing that most of the emission comes from the inner part of the imaged region ($Y_S \cong 310$ μm ± 50 μm in Fig. 5(a)), and that brighter lasing spots are appreciable in the intensity map (e.g., at $Y_S \cong 290$ μm and 340 μm). The overall spectrum from the excited nanofibers is obtained by the line-to-line sum of the intensity data from Fig. 5(a), performed for each single wavelength. This is shown in Fig. 5(b), where we highlight three sample wavelengths ($\lambda_\alpha$, $\lambda_\beta$, $\lambda_\gamma$) whose optical mode spatial profiles will be considered in the following. We then collect the emission from many vertical sections of the entire excited region, according to a detailed SSCC method [5]. This is carried out by translating the collecting lens (L3 in the scheme of Fig. 1) along the direction, *X*, perpendicular to *Y* and to the direction of the laser beam propagation (*Z*). The optics involved provide a 5× magnification of the collected emission region, thus imaging a segment with 20 μm width along $X_S$ for each 100 μm-translation of the L3 lens in the real space. Given the excitation spot size at the sample surface, an overall picture of the emission is obtained by separately recording 15 sections, each associated to a different position of the L3 lens.

Summing up the resulting individual images over the CCD vertical coordinate (*Y*), a further emission map is built up where the contributions of single sections are composed along the *X* direction as a function of wavelength (Fig. 5(c)). Here $X_S$ is the $Y_S$-perpendicular coordinate on the sample. The map evidences that most of the random lasing peaks are distributed along a few tens of μm along $X_S$, which is of the order of the delocalization measured for conjugated polymer films [5].

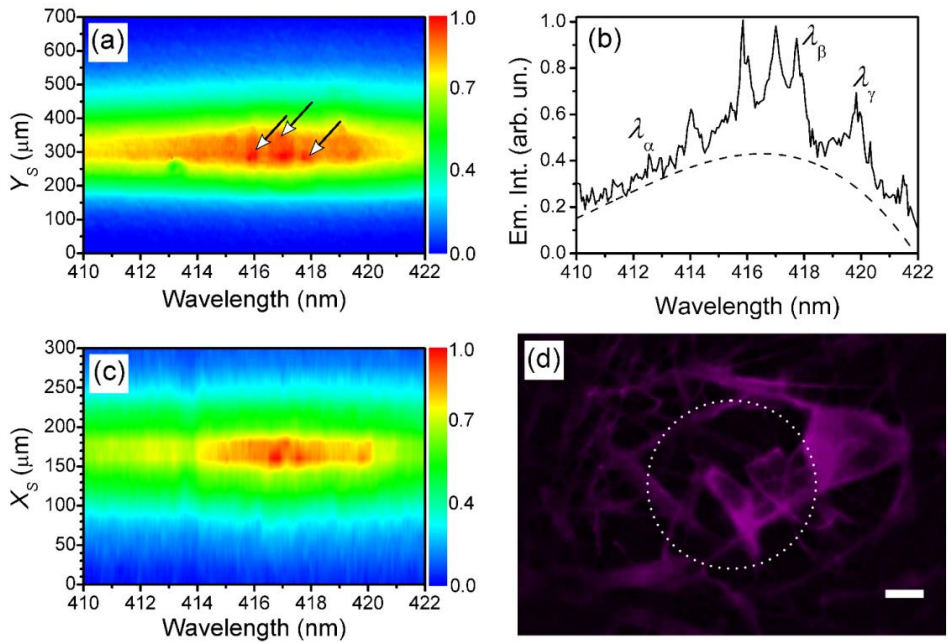

Fig. 5. (a) CCD image showing an intensity map for the spectrally and spatially($Y_S$)-resolved emission for the nanofiber random laser. $Y_S$ is the sample coordinate parallel to the long axis of the monochromator slit. The arrows indicate brighter lasing regions. (b) Intensity profile, obtained integrating the image in (a) over all the $Y$-positions for each wavelength. $\lambda_\alpha$ = 412.5 nm, $\lambda_\beta$ = 417.7 nm, and $\lambda_\gamma$ = 419.9 nm indicate three exemplary wavelengths used for the subsequent detailed SSCC analysis. Dashed line: eventually amplified spontaneous emission signal. (c) Overall, spectrally and spatially($X_S$)-resolved random lasing mode pattern, calculated summing up over the $Y$ direction each image obtained for a given position of the collection lens, and composing sub-images along the $X$ direction. Intensity data in (a-c) are averaged over 10 excitation shots. (d) Far-field fluorescence pattern recorded simultaneously with the map in (a), for the position $X_S$ = 160 µm in (c). Scale bar: 50 µm. The dashed circle highlights the ~ 250 µm sample region, which is excited directly. Excitation fluence: 42 mJ cm$^{-2}$.

Far-field fluorescence micrographs highlight unique features of random lasing nanofibers (Fig. 5(d)). Following patterns much more complex than in clusters of nanoparticles [2,34], in random media with bubble structure [35] and in conjugated polymer films [5], emission is not found to occur from all the illuminated field of the material, but instead only from the regions where multiple scattering and waveguiding in the intricate fiber networks combine with an effective interplay. The image shows a combination of different linear segments for the various modes, given by the nanofiber geometry underneath. This ramified pattern suggests the presence of different random cavities in competition in the electrospun mat, and that emission from the light-emitting fibers is also detected well far away from the directly excited area, due to waveguiding of light along the organic filaments. In Fig. 5(d), the bright filaments are clearly visible due to outcoupling of waveguided photons from the lateral surface of fiber bodies, which is supported by Rayleigh scattering from surface roughness or embedded nanoparticles. As opposed to breaks at the tips of nanofiber waveguides [29,36], here the scatterers might redistribute a fraction of radiation also forwardly, namely along the longitudinal axis of the organic filaments. The self-absorption of the gain molecule, potentially hindering waveguiding, is below $5 \times 10^2$ cm$^{-1}$ in our fibers due to the significant Stokes shift of Fl-Cz-Fl (about 0.55 eV). In principle, scattering of the incident UV excitation light by the electrospun fibers could also contribute to the broadening of the excitation laser spot when it impinges onto the fibrous sample [35]. However, given the micrometer size of the fibers (2-5 μm), which is an order of magnitude larger than the excitation wavelength (0.355 μm), such effect is not expected to be dominant in our material. For instance, calculating the angular dependence of the light scattering form factor, $f(\theta)$, for cylindrical polymer fibers [37], evidences that for $ka \gg 1$ ($a$ is the fiber radius and $k=2\pi/\lambda$ is the wavevector of the incident light) most of the incident light is scattered at forward angles < 60°. Instead, light scattering at angles around 90° (i.e. within the plane of the fiber network), which can contribute to spatial broadening of the excitation spot on the fibrous samples, becomes relevant for $ka \approx 1$, namely for fiber with size below 100 nm.

Integrating each ($\lambda$, $Y_S$) image (as that in Fig. 5(a)) over the whole width of the CCD (i.e., over wavelengths), and composing the resulting profiles obtained for each $X$-coordinate, one obtains an ($X_S$, $Y_S$) emission intensity map from the sample surface.

The spatial distribution of the overall emitted intensity is reconstructed in Fig. 6(a), where the entirety of modes (i.e., at all wavelengths and across the whole emitting region) is compelled. Isolated cavities cannot be distinguished in Fig. 6(a), suggesting significant overlap of optical modes, namely cavity density such to overcome our spatial resolution [20 μm ($X_S$)× 10 μm ($Y_S$)]. Upon restricting the spectral integration in ($\lambda$, $Y_S$) maps in the range of interest for individual modes highlighted in Fig. 5(b) (i.e. by summing intensities only for $\lambda_i \pm 0.4$ nm, where $i = \alpha, \beta, \gamma$) before composing the corresponding ($X_S$, $Y_S$)$_i$ maps, the modes at each wavelength can be finely tracked in their spatial distribution (Fig.s 6(b)-6(d)).

Here, different profiles are found, corresponding to differently shaped cavities, different degrees of delocalization within the excited region, and possibly different coupling extent for resonances across the complex material. The localization length of the mode at $\lambda_i$, $\ell_{loc-\lambda i}$, defined as the length of the portion of the system in which the amplitude of the optical state differs appreciably from zero, can be estimated from each mode map, by the inverse participation ratio, $IPR_{\lambda i}$ (how many sites the optical state is distributed over), as [10,38]:

$$\ell_{loc-\lambda i} \sim \frac{1}{\sqrt{IPR_{\lambda i}}} = \frac{\int I_{\lambda i}(x,y)dxdy}{[\int I_{\lambda i}(x,y)^2 dxdy]^{1/2}} \tag{1}$$

where $I_{\lambda i}(x,y)$ is the $\lambda_i$-mode intensity. $IPR_{\lambda i}$ are found to be below $10^{-4}$ μm$^{-2}$, indicating spread states. The resulting localization lengths (155 μm, 123 μm, and 148 μm for $\lambda_\alpha$, $\lambda_\beta$, and $\lambda_\gamma$, respectively), much larger than in cluster of nanoparticles [4,10], lead to the conclusion that modes broadly extend over the network of fibers. The overall spatial broadening of these modes, partially overlapping in space, is of the order of magnitude of the size of the excited region in our experiments. The large values of $\ell_{loc-\lambda i}$ also suggest that the number of cavities supporting lasing, that can be directly activated at a given wavelength, is quite reduced [39]. Estimated quality factors ($\geq 10^3$) and localization lengths, being of the same order for the various analyzed peaks, indicate that there is no clearly dominant optical mode. Interestingly, slightly more extended modes (such as those corresponding to $\lambda_\alpha$ and $\lambda_\gamma$) are found at spectral positions where the optical gain is lower than the maximum (expected at about $\lambda_\beta$, that is in the central region of the gain curve from Fl-Cz-Fl/PS nanofibers). This result, seemingly contradicting previous findings on modes with different extent coexisting in lasers made by fields of ZnO nanoparticles [4], can be rationalized by taking into account the extra contribution of fibers to spectral selection, with some wavelengths possibly being better supported by waveguiding and formation of loops with relatively lower optical losses [34]. Hence, the corresponding modes might benefit from a higher lifetime, thus lasing from spectral regions where a relatively lower gain is measured for Fl-Cz-Fl/PS nanofibers. Overall, electrospun non-wovens are likely to sustain mode coupling and non-locality in the fiber random laser, selecting specific wavelengths from a background of uncorrelated and spatially-decoupled modes, which could be attributed to internal waveguiding along fibers [36].

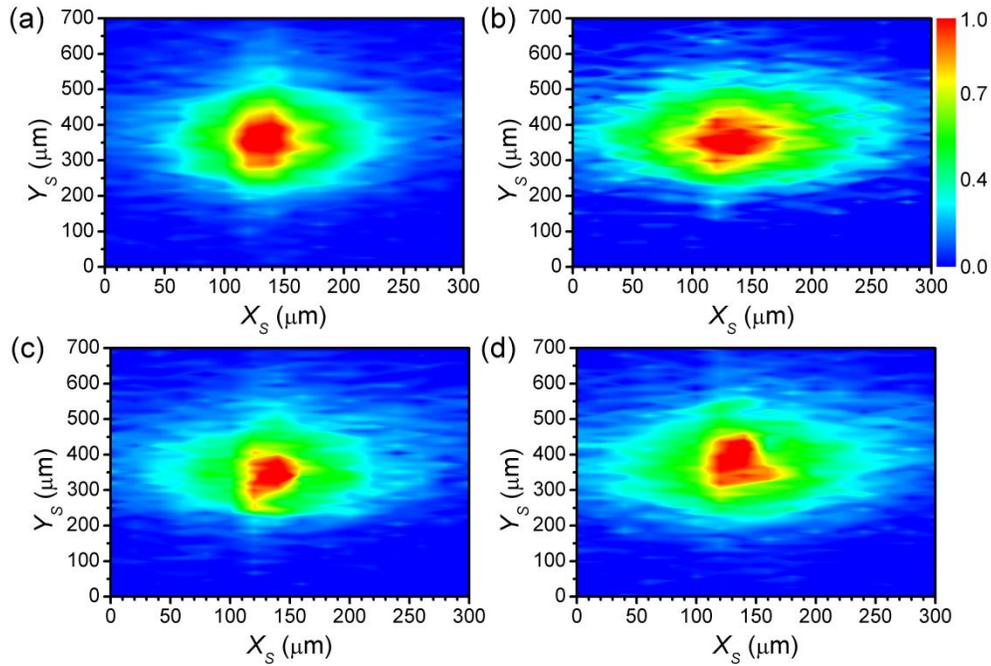

Fig. 6. (a) Spatial distribution of overall set of random lasing modes in the excitation region. (b-d) Spatial distributions of modes for $\lambda_\alpha$ (b), $\lambda_\beta$ (c) and $\lambda_\gamma$ (d), respectively. Excitation fluence: 42 mJ cm$^{-2}$. Intensity data are averaged over 10 excitation shots.

## 4. Conclusions

In conclusion, we show electrospun fibers with optical gain and light-scattering properties, which features random lasing with sub-nm spectral width as well as modes with very high spatial extent, up to the 100 μm-scale. The emission patterns allow in principle the guiding behavior of the fibers to be linked with their spectral features, active cavities, and morphology. This is an interesting outlook for future experiments.

Many applications might be opened by coupling modes from random lasers, possibly with different degrees of localization, with other sites of the same complex material or with external receivers through efficient transmission channels as provided by integrated nanofibers. Platforms especially benefiting from such architectures would include chemical and biochemical sensors, which would exploit electrospun non-wovens with ultra-high surface-to-volume ratio to detect agents affecting random lasing emission, and promptly transmit information from extended optical modes to coupled detectors. Also, fibers could be used as input components triggering emission, by altering the oscillation frequency of spatially-defined regions in the disordered material [8] in a controlled way.

## 5. Funding

The research leading to these results has received funding from the European Research Council under the European Union's Seventh Framework Programme (FP/2007-2013)/ERC Grant Agreement n. 306357 (ERC Starting Grant "NANO-JETS"). The Apulia Network of Public Research Laboratories WAFITECH (09) is also acknowledged. K.K. and S.J. acknowledge funding by a grant (No. LJB-3/2015) from the Research Council of Lithuania.

Published in Optics Express, **25**(20) 24604-24614. Doi: [10.1364/OE.25.024604](10.1364/OE.25.024604) (2017).14. B.H. Hokr, J.N. Bixler, M.T. Cone, J.D. Mason, H.T. Beier, G.D. Noojin, G.I. Petrov, L.A. Golovan, R.J. Thomas, B.A. Rockwell, and V.V. Yakovlev, "Bright emission from a random Raman laser," Nat. Commun. **5**,4356 (2014).

15. S. Gottardo, R, Sapienza, P.D. García, A. Blanco, D.S. Wiersma, and C. López, "Resonance-driven random lasing," Nat. Photonics **2**, 429-432 (2008).

16. A. Camposeo, M. Polo, P. Del Carro, L. Silvestri, S. Tavazzi, and D. Pisignano, "Random lasing in an organic light-emitting crystal and its interplay with vertical cavity feedback," Laser Photon. Rev. **2** 8, 785-791 (2014).

17. J. Gierschner, S. Varghese, and S.Y. Park, "Organic single crystal lasers," Adv. Opt. Mater. **4**, 348-364 (2016).

18. F. Quochi, "Random lasers based on organic epitaxial nanofibers," J. Opt. **12**, 024003 (2010).

19. A. Camposeo, P. Del Carro, L. Persano, K. Cyprych, A. Szukalski, L. Sznitko, J. Mysliwiec, and D. Pisignano, "Physically transient photonics: random versus distributed feedback lasing based on nanoimprinted DNA," ACS Nano **8**, 10893-10898 (2014).

20. S. Caixeiro, M. Gaio, B. Marelli, F.G. Omenetto, and R. Sapienza, "Random lasing: silk-based biocompatible random lasing," Adv. Opt. Mat. **4**, 998-1003 (2016).

21. L. Sznitko, J. Mysliwiec, and A. Miniewicz, "The role of polymers in random lasing," J. Polym. Sci., Part B: Polym. Phys. **53**, 951-974 (2015).

22. C. Sanchez, B. Lebeau, F. Chaput, and J.P. Boilot, "Optical properties of functional hybrid organic-inorganic nanocomposites," Adv. Mater. **15**, 1969-1994 (2003).
19